\documentclass[useAMS,usenatbib]{mn2e}
\usepackage{times}
\usepackage{graphicx}
\usepackage{epsfig}
\usepackage[T1]{fontenc}
\usepackage{ifpdf}
\title[Chandra Identification of Two IGR AGN]{
{\em Chandra} Identification of Two AGN Discovered by {\em INTEGRAL}}

\author[Tomsick et al.]{
John A. Tomsick$^{1}$\thanks{E-mail: jtomsick@ssl.berkeley.edu (JAT)}, 
Roman Krivonos$^{1,2}$, 
Farid Rahoui$^{3,4}$,
Marco Ajello$^{5}$, 
Jerome Rodriguez$^{6}$, \and
Nicolas Barriere$^{1}$, 
Arash Bodaghee$^{7}$, and
Sylvain Chaty$^{6,8}$
\\
$^{1}$Space Sciences Laboratory, 7 Gauss Way, University of 
California, Berkeley, CA 94720-7450, USA\\
$^{2}$Space Research Institute, Russian Academy of Sciences, 
Profsoyuznaya 84/32, 117997 Moscow, Russia\\
$^{3}$European Southern Observatory, Karl Schwarzschild-Strasse
2, 85748 Garching bei Munchen, Germany\\
$^{4}$Department of Astronomy, Harvard University, 60 Garden Street, 
Cambridge, MA 02138, USA\\
$^{5}$Department of Physics and Astronomy, Clemson University, 
Clemson, SC 29634, USA\\
$^{6}$Laboratorie AIM (UMR-E 9005 CEA/DSM-CNRS-Universit\'{e} 
Paris Diderot), Irfu/Service d'Astrophysique, CEA-Saclay, 
F-91191 Gif-sur-Yvette Cedex, France\\
$^{7}$Georgia College \& State University, CBX 82, Milledgeville,
GA 31061, USA\\
$^{8}$Institut Universitaire de France, 103 Boulevard Saint-Michel, 
75005 Paris, France}

\begin{document}


\def\lsim{\mathrel{\lower .85ex\hbox{\rlap{$\sim$}\raise
.95ex\hbox{$<$} }}}
\def\gsim{\mathrel{\lower .80ex\hbox{\rlap{$\sim$}\raise
.90ex\hbox{$>$} }}}

\pagerange{\pageref{firstpage}--\pageref{lastpage}} \pubyear{2014}

\maketitle

\label{firstpage}

\begin{abstract}

Here, we report on observations of two hard X-ray sources that were 
originally discovered with the {\em INTEGRAL} satellite: IGR~J04059+5416 
and IGR~J08297--4250.  We use the {\em Chandra X-ray Observatory} to 
localize the sources and then archival near-IR images to identify the 
counterparts.  Both sources have counterparts in the catalog of extended
2 Micron All-Sky Survey sources, and the counterpart to IGR~J04059+5416
has been previously identified as a galaxy.  Thus, we place IGR~J04059+5416 
in the class of Active Galactic Nuclei (AGN), and we suggest that 
IGR~J08297--4250 is also an AGN.  If this identification is correct, the 
near-IR images suggest that the host galaxy of IGR~J08297--4250 may be 
merging with a smaller nearby galaxy.  For IGR~J04059+5416, the 0.3--86\,keV 
spectrum from {\em Chandra} and {\em INTEGRAL} is consistent with an 
absorbed power-law with a column density of 
$N_{\rm H} = (3.1^{+2.0}_{-1.5})\times 10^{22}$\,cm$^{-2}$ and a photon index 
of $\Gamma = 1.4\pm 0.7$, and we suggest that it is a Seyfert galaxy.  
For IGR~J08297--4250, the photon index is similar, $\Gamma = 1.5\pm 0.8$, 
but the source is highly absorbed ($N_{\rm H} = (6.1^{+10.1}_{-4.3})\times 10^{23}$\,cm$^{-2}$).

\end{abstract}


\section{Introduction}

The hard X-ray imaging by the {\em INTErnational Gamma-Ray Astrophysics 
Laboratory (INTEGRAL)} satellite \citep{winkler03} has led to the discovery 
of a large number of new or previously poorly studied ``IGR'' sources.  The 
most recent published catalogs of 20--100\,keV sources detected by the Imager 
on-Board the {\em INTEGRAL} Satellite \citep[IBIS;][]{ubertini03} include more 
than 700 sources for the whole sky \citep{bird10} and 402 sources within $17.5^{\circ}$ 
of the Galactic plane\footnote{see http://hea.iki.rssi.ru/integral} \citep{krivonos12}.  
The current {\em INTEGRAL} source catalog\footnote{see http://isdc.unige.ch/integral/catalog/38/catalog.html}
includes 954 sources detected by IBIS.  The nature of $\approx$200 of 
the IGR sources is still unknown according to the most up-to-date 
list\footnote{see http://irfu.cea.fr/Sap/IGR-Sources/}.

While {\em INTEGRAL} has found many new sources, it only localizes them
to $1^{\prime}$--$5^{\prime}$, which is not nearly adequate for finding 
optical/IR counterparts.  Short exposures of IGR sources with the 
{\em Chandra X-ray Observatory} allow for a major advance in understanding 
the nature of these sources by providing sub-arcsecond positions, leading 
to unique optical/IR counterparts, as well as 0.3--10\,keV spectra that 
can be used to measure column densities and continuum shapes 
\citep[see][for examples of our previous work on {\em Chandra} follow-up
of IGR sources]{tomsick06,tomsick08,tomsick09a,tomsick12a,bodaghee12a}.  
Other groups have been carrying out similar investigations
\citep[e.g.,][]{fiocchi10,ratti10,paizis11,paizis12,nowak12,karasev12}.

Prior to this study, little was known about IGR~J04059+5416 and 
IGR~J08297--4250.  They were discovered in the \cite{krivonos12} hard 
X-ray survey, which includes 9-years of {\em INTEGRAL} data.  They are 
listed as being unidentified in the \cite{krivonos12} catalog, and there 
have not been any more publications on these sources.  A search of the 
SIMBAD database does not suggest any likely counterparts.  The {\em Chandra} 
observations that we made to identify the sources are described in Section 2.
Then, the results are presented in Section 3, including {\em Chandra}
localizations, near-IR identifications, and an analysis of the 
{\em Chandra} and {\em INTEGRAL} energy spectra.  Section 4 
includes a discussion of the results and the conclusions.

\section[]{Observations and Data Reduction}

IGR~J04059+5416 and IGR~J08297--4250 were both observed with {\em Chandra}
for 4.9\,ks in late-2013 (see Table~\ref{tab:obs}).  We used the Advanced 
CCD Imaging Spectrometer \citep[ACIS,][]{garmire03} instrument, which has
a 0.3--10\,keV bandpass, and includes two arrays of CCDs: ACIS-I and ACIS-S.  
Our observations used ACIS-I, giving a $16.9\times 16.9$\,arcmin$^{2}$ 
field-of-view (FOV), which easily covers the {\em INTEGRAL} error circles.
We obtained the data from the {\em Chandra} X-ray Center and processed it
using the {\em Chandra} Interactive Analysis of Observations (CIAO) v4.6
software and the Calibration Data Base (CALDB) v4.6.2.  We used 
{\ttfamily chandra\_repro} to produce event lists that were used for
further analysis.

\section{Results}

\subsection{{\em Chandra} Counterpart Identifications}

We searched for {\em Chandra} sources by running {\ttfamily wavdetect}
\citep{freeman02} on the ACIS-I images.  We used the same method
described in \cite{tomsick12a}, including binning the images to four
different pixel sizes.  In addition, we applied {\ttfamily wavdetect} 
to images in three different energy bands: 0.3--10\,keV, 0.3--2\,keV, 
and 2--10\,keV.  Considering the different binnings and different energy 
bands, we obtained 12 source lists, which we merged into a single list
for each observation.  We considered all sources detected at a significance
level of 2-$\sigma$ or greater.

For IGR~J04059+5416, we detect 12 sources in the ACIS-I field, and
they are listed in Table~\ref{tab:sourcelist1}.  The ACIS counts are 
determined using circular extraction regions with radii increasing 
with off-axis angle ($\theta$) in the same manner as described in
\cite{tomsick12a}.  For the position uncertainties, we add the systematic
and statistical errors in quadrature.  For all sources, the systematic
pointing error is $0.\!^{\prime\prime}64$ at 90\% confidence and $1^{\prime\prime}$ 
at 99\% confidence \citep{weisskopf05}.  The statistical uncertainty depends
on the number of counts and the off-axis angle, and we use Equation 5 from
\cite{hong05} to calculate this quantity.  The \cite{hong05} equation gives
a 95\% confidence error, but we assume a normal distribution and convert it
to a 90\% confidence error before adding it in quadrature with the 90\%
confidence systematic error.  Table~\ref{tab:sourcelist1} also includes
an estimate of the hardness of each source.  As detailed in 
Table~\ref{tab:sourcelist1} and \cite{tomsick12a}, it is defined so that 
the maximum hardness is 1.0 (all the counts in the 2--10\,keV bin) and the
minimum hardness is --1.0 (all the counts in the 0.3--2\,keV bin).

\begin{figure*}
\begin{center}
\includegraphics[width=15cm]{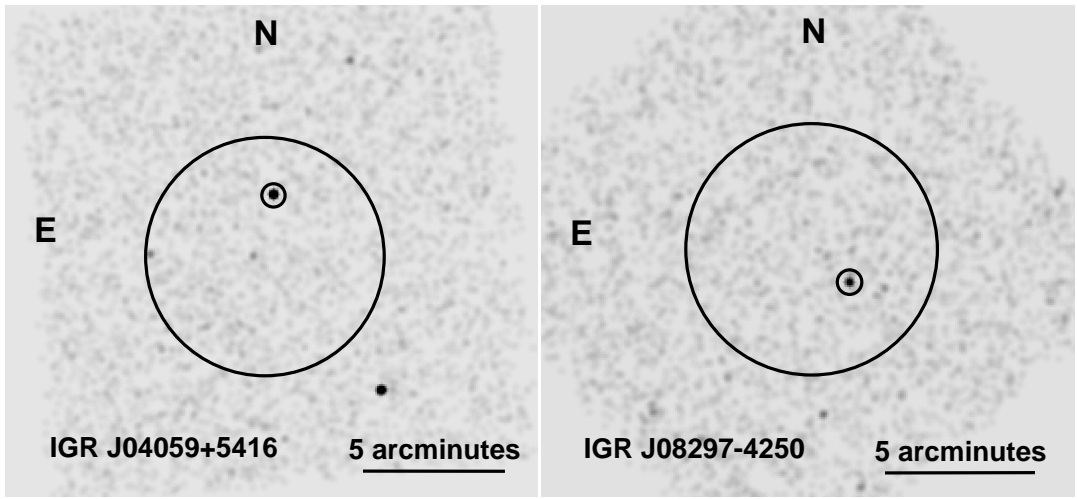}
\caption{\small {\em Chandra}/ACIS-I images in the 2--10\,keV band
for IGR~J04059+5416 {\it (left)} and IGR~J08297--4250 {\it (right)}.
In each image, the larger circle has a $4.\!^{\prime}2$ radius and 
is the 2-$\sigma$ {\em INTEGRAL} error circle.  The smaller circles
indicate the locations of the {\em Chandra} counterparts of the two
IGR sources: CXOU~J040557.6+541845 and CXOU~J082941.0--425158.  The
images have been binned and smoothed.
\label{fig:chandra_images}}
\end{center}
\end{figure*}

Source \#2 has the highest count rate, is only $2.2^{\prime}$ from the 
{\em INTEGRAL} position of IGR~J04056+5416, and has a hard spectrum.  
Source \#1 is closer to the {\em INTEGRAL} position, but Source \#2 is 
well within the 2-$\sigma$ {\em INTEGRAL} error circle (see 
Figure~\ref{fig:chandra_images}).  Considering that the count rate
for Source \#1 is an order of magnitude lower, we argue that Source \#2 is 
a much more likely counterpart.  The 2--10\,keV image shown in 
Figure~\ref{fig:chandra_images} also shows a relatively bright and hard source
to the southwest of the {\em INTEGRAL} position.  This is Source \#7, and its
count rate and hardness are only slightly lower than Source \#2.  Although
Source \#7 was not previously known to be an X-ray source, a catalog search 
shows that it is positionally coincident with a known radio source, 4C~54.04.
However, it is $6.2^{\prime}$ away from the {\em INTEGRAL} position, which is 
well outside the 2-$\sigma$ error circle and close to the 3-$\sigma$ error 
circle.  Overall, CXOU~J040557.6+541845 (Source \#2) is, by far, the most 
likely counterpart to IGR~J04059+5416, and its {\em Chandra} position is
given in Table~\ref{tab:sourcelist1}.  The 0.3--10\,keV ACIS-I count rate 
for CXOU~J040557.6+541845 is $1.89\times 10^{-2}$\,s$^{-1}$ (92.6 counts 
during the 4910\,s observation).

We produced a source list for the IGR~J08297--4250 field in the same 
manner as described above, and we find 18 sources detected by ACIS-I 
(see Table~\ref{tab:sourcelist2}).  Although Source \#5 has the highest
0.3--10\,keV count rate, it is a very soft source with a hardness of
--$0.72\pm 0.24$ (1-$\sigma$ error).  This is inconsistent with the hard
spectrum required to explain the hard X-ray flux measured by {\em INTEGRAL}.
Source \#2 is closer to the
center of the {\em INTEGRAL} error circle for IGR~J08297--4250, and 
it is a much harder source with a hardness of $>$0.6.  In fact, of the
22.2 detected photons for Source \#2, only one is in the 0.3--2\,keV
band.  Other than Source \#5, there are no sources with more than 20
counts, and the other sources with more than 5 counts listed in 
Table~\ref{tab:sourcelist2} are farther from the {\em INTEGRAL} 
position.  The fact that CXOU~J082941.0--425158 (Source \#2) is the
most likely counterpart for IGR~J08297--4250 is illustrated in
Figure~\ref{fig:chandra_images} since it is clearly the brightest
2--10\,keV source in the field.  Its {\em Chandra} position is given
in Table~\ref{tab:sourcelist2}, and the 0.3--10\,keV ACIS-I count rate 
for CXOU~J082941.0--425158 is $4.52\times10^{-3}$\,s$^{-1}$ (22.2 counts 
during the 4909\,s observation).  Below, we also discuss one of the 
fainter sources (Source \#1 with 4.2 counts), CXOU~J082940.7--425143, 
which is $16^{\prime\prime}$ from CXOU~J082941.0--425158.

\subsection{Near-IR identifications}

\begin{figure*}
\begin{center}
\includegraphics[width=10cm]{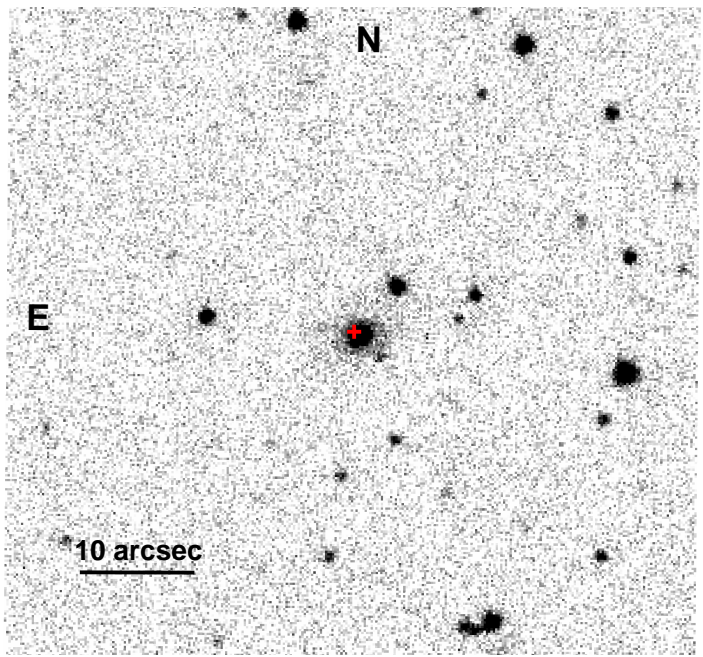}
\caption{\small UKIDSS $K$-band image for IGR~J04059+5416.  The 
cross marks the {\em Chandra} position of CXOU~J040557.6+541845, which we 
identify with 2MASX~J04055765+5418446.
\label{fig:j04059_kband}}
\end{center}
\end{figure*}

We searched the VizieR database for counterparts to the {\em Chandra}
sources at other wavelengths.  Here, we only discuss the matches that
led to conclusions about the nature of the sources.  CXOU~J040557.6+541845 
(=IGR~J04059+5416) is within $0.\!^{\prime\prime}46$ of 
2MASX~J04055765+5418446, which is from the catalog of extended sources
found in the 2MASS survey \citep{skrutskie06}.  The magnitudes are 
$J = 15.3\pm 0.2$, $H = 14.2\pm 0.2$, and $K_{s} = 13.6\pm 0.2$.  
Figure~\ref{fig:j04059_kband} shows the UKIDSS $K$-band image for 
IGR~J04059+5416, and the extended emission around 2MASX~J04055765+5418446
is visible.  The fact that this is an extended near-IR source provides 
evidence that it is an AGN.  In addition, the source is present in the 
Sloan Digitized Sky Survey, SDSS~J040557.62+541844.8, and is identified 
as a galaxy.  A third identification is with WISE~J040557.61+541844.9, 
which shows that the source is bright in the mid-IR with a 22.1\,$\mu$m 
magnitude of $6.67\pm 0.06$.  The WISE source has IR colors  
[3.4$\mu$m]-[4.6$\mu$m]=0.8 and [4.6$\mu$m]-[12$\mu$m]=2.8. These are 
typical colors of Flat Spectrum Radio Quasars (FSRQs) and Seyfert 
galaxies \citep[see, e.g.,][]{massaro2011}. FSRQs have in general a 
bright radio emission. This source does not have any associated 
counterpart in the NRAO VLA Sky Survey \citep{condon98}.  The lack of 
strong radio emission favors the Seyfert interpretation.

\begin{figure*}
\begin{center}
\includegraphics[width=15cm]{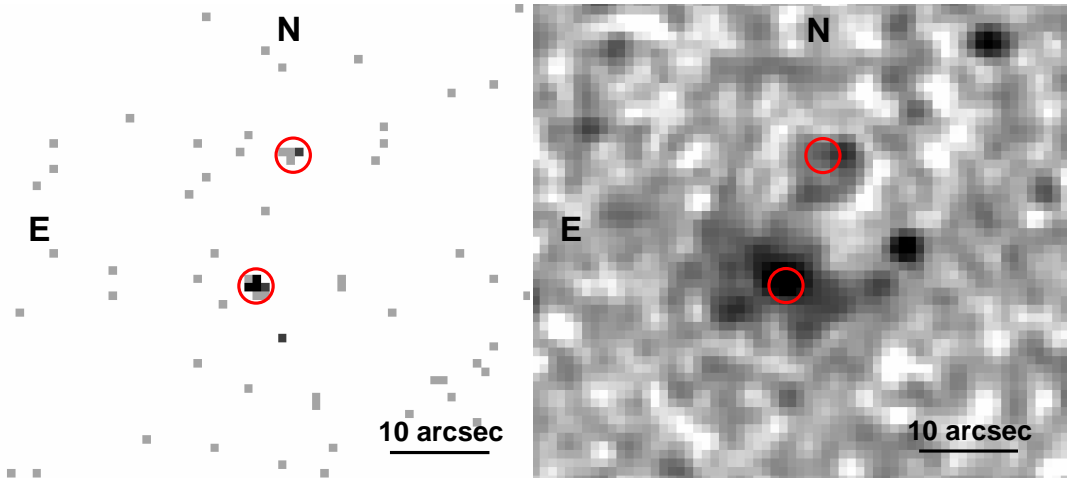}
\caption{\small {\em Chandra} 0.3--10\,keV (left) and 2MASS $K_{s}$-band (right) 
images for IGR~J08297--4250. The southern source is CXOU~J082941.0--425158
(=2MASX~J08294112--4251582), which we identify as IGR~J08297--4250.  
The northern source is CXOU~J082940.7--425143.  The centers of the red circles
are coincident with the {\em Chandra} positions.  The radii of these circles
are $2^{\prime\prime}$, but the position uncertainties are significantly smaller
($0.\!^{\prime\prime}64$ at 90\% confidence).\label{fig:j08297_images}}
\end{center}
\end{figure*}

CXOU~J082941.0--425158 (=IGR~J08297--4250) is also present in the 2MASS
catalog of extended sources being $0.\!^{\prime\prime}44$ from 
2MASX~J08294112--4251582.  Its magnitudes are $J = 14.0\pm 0.2$, 
$H = 12.5\pm 0.1$, and $K_{s} = 12.3\pm 0.1$.  Figure~\ref{fig:j08297_images}
shows extended $K_{s}$-band emission out to $\approx$10$^{\prime\prime}$.  
We also identify this source with WISE~J082941.14-425157.8, which 
has a 22.1\,$\mu$m magnitude of $6.37\pm 0.07$. 
The identification of IGR~J08297--4250 with an extended near-IR
source disfavors or rules out hard X-ray emitting Galactic populations.  
A High-Mass X-ray Binary (HMXB) might be as bright in the near-IR as 
2MASX~J08294112--4251582, but HMXBs do not show extended near-IR 
emission.  Another type of source to consider is the class of Pulsar 
Wind Nebulae (PWNe), which harbor rotation-powered pulsars.  These sources 
have extended X-ray and radio emission, but they are extremely faint in 
the optical and near-IR.  Out of 1800 rotation-powered pulsars, only 12 
have UV, optical, or near-IR counterparts, and, with the exception of
the Crab Nebula, they are all fainter than 22nd magnitude \citep{mignani11}.
Near-IR counterparts have also not been detected in observations of 
some of the brighter {\em INTEGRAL} PWNe \citep{curran11}. Thus, an AGN 
identification is very likely for IGR~J08297--4250.  

There is a second near-IR source about $16^{\prime\prime}$ to the North 
that also appears to be extended.  It does not appear in the extended 
2MASS catalog, but it is 2MASS~J08294046--4251430 with magnitudes 
$H = 16.0\pm 0.2$ and $K_{s} = 15.0\pm 0.1$.  We suggest a possible 
identification of this source with CXOU~J082940.7--425143.  The 
{\em Chandra} position falls within the extended emission, but it is 
$2.\!^{\prime\prime}6$ from the brightest part of the extended near-IR
emission.  The 2MASS astrometry is good to $\sim$$0.\!^{\prime\prime}1$, 
indicating that the {\em Chandra} source position is clearly inconsistent 
with the brightest part of the near-IR emission.  Whether the 
{\em Chandra}/2MASS association is valid or not, if 2MASS~J08294046--4251430 
is a galaxy, then it is possible that it is interacting with 
2MASX~J08294112--4251582.  Thus, we consider these sources to potentially 
be a pair of merging galaxies.

\subsection{Energy Spectra}

We produced 0.3--10\,keV {\em Chandra} source and background energy 
spectra and response files using the CIAO script {\ttfamily specextract}.  
We used a circular source extraction region with a radius of 5 pixels 
($2.\!^{\prime\prime}5$) and obtained a background spectrum from a nearby 
source-free region.  For part of our analysis, we also use 17--86\,keV 
energy spectra from {\em INTEGRAL}/IBIS.  The {\em INTEGRAL} spectra
are data products obtained as part of the \cite{krivonos12} study, and
they represent average spectra over a 9-year period.  For IGR~J04059+5416 
and IGR~J08297--4250, the {\em INTEGRAL} spectra include exposure times 
of 1.3\,Ms and 5.6\,Ms, respectively.  In the following,
we first perform spectral fitting for the two IGR sources with 
{\em Chandra} only, and then we jointly fit the {\em Chandra} and 
{\em INTEGRAL} spectra.  Some caution is necessary regarding the joint
fits since we are combining a single {\em Chandra} snapshot with 
{\em INTEGRAL}'s 9-year average.  All spectral fitting is done with 
XSPEC v12.8.2.  

\subsubsection{IGR~J04059+5416}

Due to the low number of counts, we fit the {\em Chandra} spectra by
minimizing the Cash statistic \citep{cash79}.  Although this statistic does 
not require binning of the data, we made a spectrum with a signal-to-noise
ratio of at least 3 in each bin (except for the highest energy bin where
the signal-to-noise ratio is 2.8) in order to look for any significant 
features in the residuals and to use the $\chi^{2}$ value as a representative 
determination of the quality of the fit.  A simple absorbed power-law model 
provides an acceptable fit with a Cash statistic of 2.0 and a reduced-$\chi^{2}$ 
of 0.35 for 5 degrees of freedom (dof).  As shown in Table~\ref{tab:parameters04}, 
the column density, which is calculated using \cite{wam00} abundances and 
\cite{vern96} cross sections, is $N_{\rm H} = (3.5^{+2.4}_{-1.9})\times 10^{22}$~cm$^{-2}$
(90\% confidence errors are given here and throughout the paper).  This 
is significantly higher than the Galactic value along the IGR~J04059+5416 
line of sight, $N_{\rm H} = 7\times 10^{21}$\,cm$^{-2}$ \citep{kalberla05}, 
indicating that some of the absorption is from the AGN's host galaxy.  The 
photon index from the {\em Chandra}-only fit is $\Gamma = 1.5^{+0.9}_{-0.8}$.

We also performed spectral fits to the {\em Chandra}+{\em INTEGRAL}
joint spectrum (see Table~\ref{tab:parameters04} and Figure~\ref{fig:spectra}).  
We used $\chi^{2}$ minimization due to the high level of background for 
{\em INTEGRAL}.  As before, we fit with an absorbed power-law model, and we 
also included a constant parameter, which allows for different overall 
normalizations for the two spectra (due to, e.g., possible source variability).  
For the joint fit, there is very little change to the $N_{\rm H}$ and $\Gamma$ 
parameters.  With a value of $2.9^{+8.9}_{-2.2}$, the {\em INTEGRAL} normalization 
relative to {\em Chandra} does not allow us to draw conclusions on the source 
variability (1.0 indicating no variability).  The constant source case would 
imply a harder spectrum (see Table~\ref{tab:parameters04}).  For all the 
IGR~J04059+5416 fits, we only included the three lowest energy {\em INTEGRAL} 
bins in the fit because the source was not detected in the highest energy bin.  
The 2-$\sigma$ upper limit on the flux in the the 57--86\,keV bin is somewhat 
lower than the power-law (see Figure~\ref{fig:spectra}), which could be the 
sign of a cutoff.

\subsubsection{IGR~J08297--4250}

\begin{figure*}
\begin{center}
\includegraphics[width=15cm]{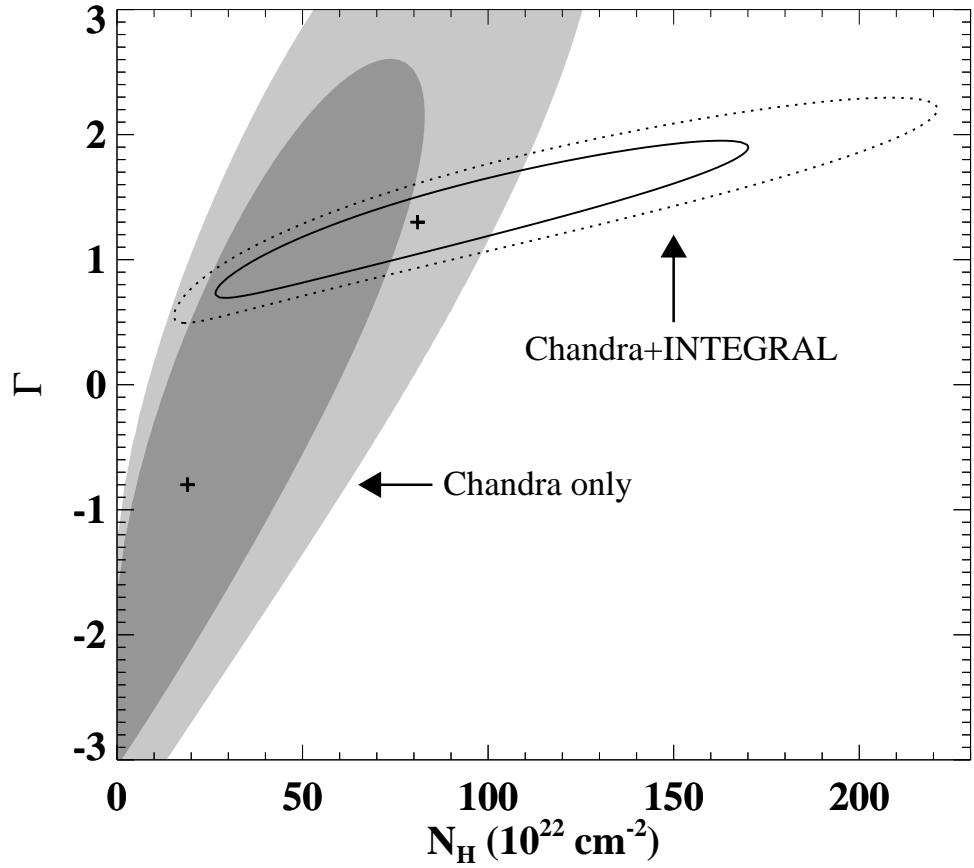}
\caption{\small Contour plot for absorbed power-law fits to the IGR~J08297--4250 spectra.  
The parameters are the column density ($N_{\rm H}$) and the power-law index ($\Gamma$).  
For the fits to the {\em Chandra} spectrum, the dark shaded region corresponds to the 
1-$\sigma$ error range ($\Delta$$\chi^{2} = 2.3$), and the light shaded region corresponds
to the 90\% confidence error range ($\Delta$$\chi^{2} = 4.6$).  For the joint fits to the
{\em Chandra} and {\em INTEGRAL} spectrum, the solid line encloses the 1-$\sigma$ error
range, and the dotted line encloses the 90\% confidence error range.\label{fig:contour}}
\end{center}
\end{figure*}

Here, we also began by fitting just the {\em Chandra} spectrum with an
absorbed power-law model and minimizing the Cash statistic.  This 
provides an acceptable fit with a Cash statistic of 6.4 and a 
reduced-$\chi^{2}$ of 0.65 for 5 dof.  For binning, we used a 
signal-to-noise ratio requirement of at least 1.25 per bin.  The best
fit spectral parameters are $N_{\rm H} = 1.9\times 10^{23}$\,cm$^{-2}$
and $\Gamma$ = --0.8, but they are both very poorly constrained since
the spectrum only has 22 counts.  We produced error contours for 
$N_{\rm H}$ and $\Gamma$, and these are shown in Figure~\ref{fig:contour}.
The two parameters are highly-degenerate, and it is unclear if the
spectrum is intrinsically very hard or if the column density is very
high.

\begin{figure*}
\begin{center}
\includegraphics[width=15cm]{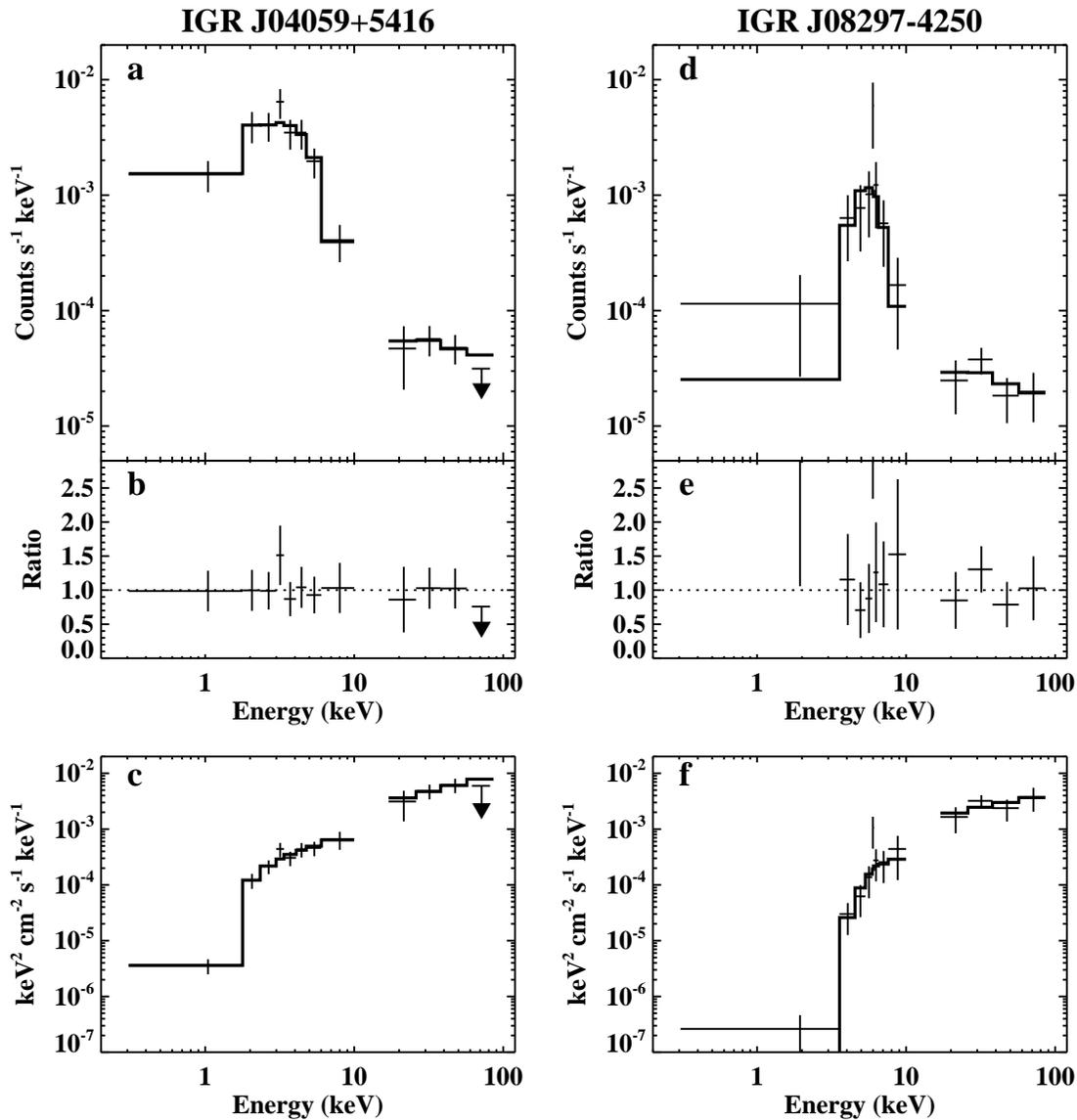}
\caption{\small {\em Chandra} and {\em INTEGRAL} spectra for IGR~J04059+5416 (3 left panels) and IGR~J08297--4250 (3 right panels).  {\it (a)} and {\it (d)}: Counts spectra fitted with an absorbed power-law model.  For IGR~J04059+5416, the 2-$\sigma$ upper limit is shown for the highest-energy {\em INTEGRAL} bin.  {\it (b)} and {\it (e)}: Data-to-model ratios.  {\it (c)} and {\it (f)}:  Unfolded spectra.  
\label{fig:spectra}}
\end{center}
\end{figure*}

As shown in Figure~\ref{fig:spectra}, the spectrum shows a high point 
at $6.0\pm 0.1$\,keV, and this could be interpreted as a noise 
fluctuation or a redshifted Fe K$\alpha$ emission line.  When a narrow 
Gaussian is added, the Cash statistic drops from 6.4 to 0.4.  To 
determine if the Gaussian is required by the data, we used 
{\ttfamily simftest} to make 1000 simulated spectra with the absorbed
power-law model as the parent distribution.  For the significance
test, we used a spectrum with 21 bins to lessen any bias due to
binning.  With the actual data, the Cash statistic decreases from 
16.7 to 10.1 for 17 dof.  When the simulated spectra are fit with 
and without a narrow Gaussian in the 5--7\,keV range, they show 
improvements by at least as much as the actual data $\approx$5\% 
of the time, and we conclude that the significance of the line is 
2.0-$\sigma$.  We consider this to be a marginal detection of a 
redshifted Fe K$\alpha$ line.

Fitting the {\em INTEGRAL} spectrum jointly with {\em Chandra} greatly 
improves the constraints on the spectral parameters 
(Table~\ref{tab:parameters08}).  A very large column density of
$N_{\rm H} = (6^{+10}_{-4})\times 10^{23}$\,cm$^{-2}$ is required, and
while {\em Chandra} alone allowed for the possibility of a very
hard power-law spectrum, we find $\Gamma = 1.5\pm 0.8$, which is
typical for AGN.  The column density is very much in excess of
the Galactic value along the line of sight.  While the {\em INTEGRAL} 
normalization relative to {\em Chandra} allows for significant 
variability, the parameters do not change significantly for 
the constant source case, and the error contours for these parameters
are shown in Figure~\ref{fig:contour}.  The addition of the {\em INTEGRAL} 
data allows us to rule out the low-$N_{\rm H}$/hard-$\Gamma$ solution 
that was a possibility with {\em Chandra} alone.

Figure~\ref{fig:spectra} shows that the simple absorbed power-law 
fit results in a mismatch between the model and the lowest energy bin
in the {\em Chandra} spectrum.  Although the descrepancy is only 
1-$\sigma$, the addition of absorption with partial covering provides
a good explanation, and the ratio residual for this bin goes from being 
off-scale in Figure~\ref{fig:spectra}e (actually, the value of the
ratio is $4.2\pm 3.2$) to being very close to 1.0 ($0.9\pm 0.7$).
While it is likely that there is partial covering absoprtion for
this source, it does not provide a large enough improvement in the
fit to be formally required, and the statistical quality of the
spectrum is not sufficient to constrain the parameters.

\section{Discussion and Conclusions}

For our larger {\em Chandra} study in 2013--2014, we chose ten 
unidentified {\em INTEGRAL} sources from the \cite{krivonos12} 
catalog.  While analysis of the data is still on-going (and 
one observation has not yet occurred), the observations that
we report on here have yielded an identification of one AGN 
(IGR~J04059+5416) and one likely AGN (IGR~J08297--4250).
Of the 402 sources in the 
\cite{krivonos12} catalog, 112 were identified as AGN or AGN 
candidates (with 104 being confirmed AGN).  Although we selected 
sources within $8^{\circ}$ of the Galactic plane (compared to the 
full catalog, which goes out to $17.5^{\circ}$), and IGR~J04059+5416 
and IGR~J08297--4250 have Galactic latitudes of $b = 1.54^{\circ}$ 
and $b$ = --$2.21^{\circ}$, respectively, it is still not extremely 
surprising for 20\% of our sample to be composed of AGN.  
In our previous studies, we have identified several {\em INTEGRAL}
sources near the Galactic Plane as AGN \citep{chaty08,zct09,tomsick12a}.

The 112 AGN or AGN candidates in the \cite{krivonos12} catalog have
17--60\,keV fluxes ranging from 0.37 mCrab to 12 mCrab with a median 
flux of 0.98 mCrab.  In the \cite{krivonos12} catalog, the 
17--60\,keV fluxes of IGR~J04059+5416 and IGR~J08297--4250 are 
0.86 and 0.35 mCrab, respectively.  Thus, IGR~J04059+5416 is close 
to the median flux, and IGR~J08297--4250 is slightly fainter
than the faintest source identified as an AGN in the catalog.

For IGR~J04059+5416, the optical and near-IR information provides
the strongest evidence that it is an AGN, but the type of AGN is 
not immediately clear.  The {\em Chandra}+{\em INTEGRAL} spectral
fit for the $N_{INTEGRAL}/N_{Chandra} = 1$ gives a very
hard spectrum ($\Gamma = 0.9\pm 0.2$), which might argue that the
source is a blazar.  However, the non-detection in the 57--86\,keV
band would require a sharp spectral cutoff, and this would not
be expected for a blazar at this energy.  It is more likely that 
the source is moderately variable and that it is a Seyfert galaxy.  
The 9-year 17--60\,keV IBIS light curve\footnote{see http://www.mpa-garching.mpg.de/integral/nine-years-galactic-survey/index.php?srcid=837} does not show statistically significant
variability, but the errors on individual flux measurements are large, 
and we calculate that the light curve allows for year-to-year flux 
changes by as much as a factor of two or three.  Thus, it is possible 
that the {\em Chandra} observation occurred when the flux was relatively 
low.

For IGR~J08297--4250, the {\em Chandra}+{\em INTEGRAL} fits indicate
a power-law with an index consistent with what would be expected for 
a Seyfert galaxy, and the high absorption may indicate a Seyfert 2
classification.  Although the 6.0\,keV emission line is only significant
at the 2-$\sigma$ level, if its rest-frame energy is 6.4\,keV
(neutral iron K$\alpha$), then the redshift is $z = 0.07$.  For
an unabsorbed 0.3--10\,keV flux of $2\times 10^{-12}$\,erg\,cm$^{-2}$\,s$^{-1}$, 
the corresponding luminosity would be $\sim$$10^{43}$\,erg\,s$^{-1}$, 
which is typical for hard X-ray selected Seyfert 2 AGN \citep{ajello12}.

\cite{koss2010} estimated that up to 25\% of the AGN detected above 15\,keV 
by {\it Swift}/BAT  are found in hosts in an apparent state of merging 
with a nearby ($<$30\,kpc) galaxy. This, together with the fact that AGN 
luminosity increases for decreasing separation of the galaxy pair 
\citep{koss2012}, highlights the role that merging has in triggering AGN 
activity. It is thus not unlikely to find an AGN hosted in a merging pair 
while following up on newly detected {\em INTEGRAL} sources. The majority of 
{\em INTEGRAL} Seyfert-like AGN are detected at z$\leq$0.1, implying that 
the separation between CXOU~J082941.0--425158 and  CXOU~J082940.7--425143
(assuming they are a merging pair of galaxies) is $<$30\,kpc. \cite{koss2012} 
found that systems with a close companion (e.g. $<$30\,kpc) often display 
AGN activity in both hosts, similarly to what we find here.  Moreover, 
some of the hosts of close merging pairs present a disrupted morphology 
\citep{koss2012}, which might explain the offset of CXOU~J082940.7--425143 
from the brightest part of the extended near-IR emission.  We thus believe 
that, on the basis of the evidence gathered so far, IGR~J08297--4250 is an 
AGN that might possibly be hosted in a merging pair. However, final 
confirmation will have to await higher resolution near-IR imaging and 
optical spectroscopy.


\begin{table*}
\caption{{\em Chandra} Observation Log\label{tab:obs}}
\begin{minipage}{\linewidth}
\begin{tabular}{ccccccc} \hline \hline
Target       & $l$ (deg.)  & $b$ (deg.) &  ObsID & Start Time (UT) & End Time (UT) & Exposure Time (s)\\ \hline\hline
IGR J04059+5416  & 148.93 & +1.54  & 15792 & 2013 Nov 22, 18.7 h & 2013 Nov 22, 20.9 h & 4910\\
IGR J08297--4250 & 261.08 & --2.21 & 15793 & 2013 Dec 4, 20.3 h & 2013 Dec 4, 22.1 h & 4909\\ \hline
\end{tabular}
\end{minipage}
\end{table*}

\begin{table*}
\caption{{\em Chandra} Sources in IGR J04059+5416 Field\label{tab:sourcelist1}}
\begin{minipage}{\linewidth}
\footnotesize
\begin{tabular}{ccccccc} \hline \hline
Source & $\theta$\footnote{The angular distance between the center of the {\em INTEGRAL} error circle, which is also the approximate {\em Chandra} aimpoint, and the source.} & {\em Chandra} R.A. & {\em Chandra} Decl. & ACIS  & Position &  \\
Number & (arcminutes) & (J2000)  & (J2000) &  Counts\footnote{The number of ACIS-I counts detected (after background subtraction) in the 0.3--10 keV band.} & Uncertainty\footnote{The 90\% confidence uncertainty on the position, including statistical and systematic contributions.} & Hardness\footnote{The hardness is given by $(C_{2}-C_{1})/(C_{2}+C_{1})$, where $C_{2}$ is the number of counts in the 2--10 keV band and $C_{1}$ is the number of counts in the 0.3--2 keV band.}\\ \hline
1  & 0.40 & $04^{\rm h}06^{\rm m}02^{\rm s}.49$ & +$54^{\circ}16^{\prime}34.\!^{\prime\prime}9$ &  8.6 & $0.\!^{\prime\prime}75$ &  $>$+0.1\\
2  & 2.21 & $04^{\rm h}05^{\rm m}57^{\rm s}.69$ & +$54^{\circ}18^{\prime}45.\!^{\prime\prime}1$ & 92.6 & $0.\!^{\prime\prime}71$ &  +$0.59\pm 0.14$\\
3  & 2.72 & $04^{\rm h}05^{\rm m}48^{\rm s}.84$ & +$54^{\circ}14^{\prime}21.\!^{\prime\prime}3$ &  8.6 & $0.\!^{\prime\prime}86$ &  $<$--0.2\\
4  & 3.64 & $04^{\rm h}06^{\rm m}14^{\rm s}.57$ & +$54^{\circ}19^{\prime}29.\!^{\prime\prime}7$ &  6.6 & $1.\!^{\prime\prime}11$ &  +$0.1\pm 0.7$\\
5  & 4.04 & $04^{\rm h}06^{\rm m}27^{\rm s}.44$ & +$54^{\circ}16^{\prime}38.\!^{\prime\prime}9$ & 14.3 & $0.\!^{\prime\prime}94$ &  +$0.4\pm 0.4$\\
6  & 5.60 & $04^{\rm h}05^{\rm m}26^{\rm s}.99$ & +$54^{\circ}13^{\prime}38.\!^{\prime\prime}6$ &  7.3 & $1.\!^{\prime\prime}85$ &  +$0.0\pm 0.6$\\
7  & 6.22 & $04^{\rm h}05^{\rm m}31^{\rm s}.83$ & +$54^{\circ}11^{\prime}51.\!^{\prime\prime}8$ & 84.3 & $0.\!^{\prime\prime}87$ &  +$0.37\pm 0.14$\\
8  & 6.44 & $04^{\rm h}05^{\rm m}54^{\rm s}.96$ & +$54^{\circ}10^{\prime}09.\!^{\prime\prime}7$ &  6.3 & $2.\!^{\prime\prime}75$ &  $>$--0.3\\
9  & 7.20 & $04^{\rm h}06^{\rm m}01^{\rm s}.35$ & +$54^{\circ}23^{\prime}45.\!^{\prime\prime}5$ &  8.2 & $2.\!^{\prime\prime}90$ &  $>$0.0\\
10 & 7.51 & $04^{\rm h}05^{\rm m}39^{\rm s}.06$ & +$54^{\circ}23^{\prime}26.\!^{\prime\prime}5$ & 15.2 & $2.\!^{\prime\prime}03$ &  +$0.5\pm 0.4$\\
11 & 7.64 & $04^{\rm h}05^{\rm m}44^{\rm s}.57$ & +$54^{\circ}23^{\prime}52.\!^{\prime\prime}1$ &  7.2 & $3.\!^{\prime\prime}76$ &  $<$+0.3\\
12 & 8.25 & $04^{\rm h}05^{\rm m}18^{\rm s}.97$ & +$54^{\circ}10^{\prime}51.\!^{\prime\prime}5$ &  5.2 & $6.\!^{\prime\prime}28$ &  $<$+0.2\\ \hline
\end{tabular}
\end{minipage}
\end{table*}

\begin{table*}
\caption{{\em Chandra} Sources in IGR J08297--4250 Field\label{tab:sourcelist2}}
\begin{minipage}{\linewidth}
\footnotesize
\begin{tabular}{ccccccc} \hline \hline
Source & $\theta$\footnote{The angular distance between the center of the {\em INTEGRAL} error circle, which is also the approximate {\em Chandra} aimpoint, and the source.} & {\em Chandra} R.A. & {\em Chandra} Decl. & ACIS  & Position &  \\
Number & (arcminutes) & (J2000)  & (J2000) &  Counts\footnote{The number of ACIS-I counts detected (after background subtraction) in the 0.3--10 keV band.} & Uncertainty\footnote{The 90\% confidence uncertainty on the position, including statistical and systematic contributions.} & Hardness\footnote{The hardness is given by $(C_{2}-C_{1})/(C_{2}+C_{1})$, where $C_{2}$ is the number of counts in the 2--10 keV band and $C_{1}$ is the number of counts in the 0.3--2 keV band.}\\ \hline
 1 & 1.58  & $08^{\rm h}29^{\rm m}40^{\rm s}.70$ & --$42^{\circ}51^{\prime}43.\!^{\prime\prime}3$ &  4.2 & $0.\!^{\prime\prime}88$ &  $>$--0.4\\
 2 & 1.67  & $08^{\rm h}29^{\rm m}41^{\rm s}.09$ & --$42^{\circ}51^{\prime}58.\!^{\prime\prime}5$ & 22.2 & $0.\!^{\prime\prime}73$ &  $>$+0.6\\
 3 & 1.79  & $08^{\rm h}29^{\rm m}57^{\rm s}.73$ & --$42^{\circ}50^{\prime}58.\!^{\prime\prime}4$ &  3.2 & $0.\!^{\prime\prime}98$ &  $>$--0.7\\
 4 & 2.42  & $08^{\rm h}29^{\rm m}46^{\rm s}.59$ & --$42^{\circ}53^{\prime}17.\!^{\prime\prime}0$ &  3.2 & $1.\!^{\prime\prime}11$ &  ---\\
 5 & 2.77  & $08^{\rm h}29^{\rm m}34^{\rm s}.65$ & --$42^{\circ}52^{\prime}10.\!^{\prime\prime}6$ & 40.2 & $0.\!^{\prime\prime}74$ & --$0.7\pm 0.2$\\
 6 & 3.59  & $08^{\rm h}29^{\rm m}36^{\rm s}.99$ & --$42^{\circ}47^{\prime}54.\!^{\prime\prime}8$ &  8.2 & $1.\!^{\prime\prime}01$ & --$0.4\pm 0.6$\\
 7 & 4.70  & $08^{\rm h}30^{\rm m}13^{\rm s}.58$ & --$42^{\circ}50^{\prime}32.\!^{\prime\prime}1$ &  5.8 & $1.\!^{\prime\prime}59$ &  $<$--0.2\\
 8 & 5.55  & $08^{\rm h}29^{\rm m}45^{\rm s}.96$ & --$42^{\circ}56^{\prime}24.\!^{\prime\prime}8$ & 13.8 & $1.\!^{\prime\prime}28$ &  +$0.4\pm 0.4$\\
 9 & 5.69  & $08^{\rm h}30^{\rm m}01^{\rm s}.78$ & --$42^{\circ}55^{\prime}59.\!^{\prime\prime}0$ &  5.8 & $2.\!^{\prime\prime}23$ &  $<$+0.2\\
10 & 6.68  & $08^{\rm h}30^{\rm m}12^{\rm s}.10$ & --$42^{\circ}45^{\prime}52.\!^{\prime\prime}3$ & 10.8 & $2.\!^{\prime\prime}00$ &  $<$--0.4\\
11 & 6.87  & $08^{\rm h}29^{\rm m}41^{\rm s}.92$ & --$42^{\circ}44^{\prime}06.\!^{\prime\prime}2$ &  2.8 & $6.\!^{\prime\prime}52$ &  ---\\
12 & 7.21  & $08^{\rm h}29^{\rm m}23^{\rm s}.60$ & --$42^{\circ}45^{\prime}13.\!^{\prime\prime}3$ &  2.9 & $7.\!^{\prime\prime}41$ &  ---\\
13 & 7.31  & $08^{\rm h}29^{\rm m}35^{\rm s}.41$ & --$42^{\circ}57^{\prime}49.\!^{\prime\prime}0$ &  4.9 & $4.\!^{\prime\prime}73$ &  $>$--0.9\\
14 & 7.32  & $08^{\rm h}30^{\rm m}08^{\rm s}.18$ & --$42^{\circ}57^{\prime}12.\!^{\prime\prime}0$ &  7.9 & $3.\!^{\prime\prime}12$ &  $>$+0.1\\
15 & 7.51  & $08^{\rm h}30^{\rm m}16^{\rm s}.72$ & --$42^{\circ}56^{\prime}14.\!^{\prime\prime}5$ & 16.9 & $1.\!^{\prime\prime}90$ &  +$0.5\pm 0.4$\\
16 & 8.11  & $08^{\rm h}29^{\rm m}39^{\rm s}.08$ & --$42^{\circ}58^{\prime}49.\!^{\prime\prime}2$ &  4.9 & $6.\!^{\prime\prime}33$ &  ---\\
17 & 8.42  & $08^{\rm h}29^{\rm m}03^{\rm s}.26$ & --$42^{\circ}48^{\prime}58.\!^{\prime\prime}3$ & 11.9 & $3.\!^{\prime\prime}17$ &  $>$+0.1\\
18 & 10.25 & $08^{\rm h}30^{\rm m}43^{\rm s}.94$ & --$42^{\circ}50^{\prime}53.\!^{\prime\prime}8$ &  4.9 & $13.\!^{\prime\prime}07$ & ---\\ \hline
\end{tabular}
\end{minipage}
\end{table*}

\begin{table*}
\caption{IGR~J04059+5416 Power-law Fit Parameters\label{tab:parameters04}}
\begin{minipage}{\linewidth}
\begin{tabular}{ccccc} \hline \hline
Parameter\footnote{The errors on the parameters are 90\% confidence.} & Units & {\em Chandra}-only & \multicolumn{2}{c}{{\em Chandra}+{\em INTEGRAL}}\\ \hline\hline
$N_{\rm H}$\footnote{The column density is calculated assuming \cite{wam00} abundances and \cite{vern96} cross sections.  Along this line of sight, the Galactic value is $N_{\rm H} = 7\times 10^{21}$\,cm$^{-2}$ \citep{kalberla05}.} & $10^{22}$\,cm$^{-2}$ &  $3.5^{+2.4}_{-1.9}$ & $3.1^{+2.0}_{-1.5}$ & $2.1^{+1.2}_{-0.9}$\\
$\Gamma$   &    --    & $1.5^{+0.9}_{-0.8}$ & $1.4\pm 0.7$ & $0.9\pm 0.2$\\
Unabs. Flux (0.3--10\,keV) & $10^{-12}$\,erg\,cm$^{-2}$\,s$^{-1}$ & $1.9^{+2.7}_{-0.6}$ & $1.8^{+1.2}_{-0.5}$ & $1.6^{+0.4}_{-0.3}$\\
Abs. Flux (0.3--10\,keV)  & $10^{-12}$\,erg\,cm$^{-2}$\,s$^{-1}$  & $1.1^{+0.3}_{-0.4}$ & $1.1^{+0.2}_{-0.4}$ & $1.3^{+0.2}_{-0.3}$\\
$N_{INTEGRAL}/N_{Chandra}$\footnote{The {\em INTEGRAL} normalization relative to {\em Chandra}.} &   --   &   --   & $2.9^{+8.9}_{-2.2}$ & 1.0\footnote{Fixed.}\\ \hline
\end{tabular}
\end{minipage}
\end{table*}

\begin{table*}
\caption{IGR~J08297--4250 Power-law Fit Parameters\label{tab:parameters08}}
\begin{minipage}{\linewidth}
\begin{tabular}{cccc} \hline \hline
Parameter\footnote{The errors on the parameters are 90\% confidence.} & Units & \multicolumn{2}{c}{{\em Chandra}+{\em INTEGRAL}}\\ \hline\hline
$N_{\rm H}$\footnote{The column density is calculated assuming \cite{wam00} abundances and \cite{vern96} cross sections.  Along this line of sight, the Galactic value is $N_{\rm H} = 9\times 10^{21}$\,cm$^{-2}$ \citep{kalberla05}.} & $10^{22}$\,cm$^{-2}$ & $61^{+101}_{-43}$ & $81^{+100}_{-56}$\\
$\Gamma$   &    --    & $1.5\pm 0.8$ & $1.3^{+0.7}_{-0.6}$\\
Unabs. Flux (0.3--10\,keV) & $10^{-12}$\,erg\,cm$^{-2}$\,s$^{-1}$ & $1.6^{+10.4}_{-1.2}$ & $2.0^{+14.0}_{-1.6}$\\
Abs. Flux (0.3--10\,keV)  & $10^{-12}$\,erg\,cm$^{-2}$\,s$^{-1}$  & $0.3^{+0.1}_{-0.2}$ & $0.3\pm 0.1$\\
$N_{INTEGRAL}/N_{Chandra}$\footnote{The {\em INTEGRAL} normalization relative to {\em Chandra}.} &   --   & $2.5^{+12.2}_{-2.0}$ & 1.0\footnote{Fixed.}\\ \hline
\end{tabular}
\end{minipage}
\end{table*}

\section*{Acknowledgments}

Support for this work was provided by the National Aeronautics and Space 
Administration (NASA) through {\em Chandra} Award Number GO4-15044X issued by
the Chandra X-ray Observatory Center, which is operated by the Smithsonian
Astrophysical Observatory for and on behalf of NASA under contract NAS8-03060.
This publication makes use of data products from the Two Micron All Sky Survey, 
which is a joint project of the University of Massachusetts and the Infrared 
Processing and Analysis Center/California Institute of Technology, funded by
the NASA and the National Science Foundation.  This research has made use 
of the VizieR catalogue access tool, the SIMBAD database (CDS, Strasbourg, 
France), and the IGR Sources page maintained by J. Rodriguez \& A. Bodaghee.  



\label{lastpage}
\bsp

\end{document}